\documentclass[journal]{IEEEtran}
\usepackage{cite}
\usepackage{amsmath} 

\usepackage{subfigure}
\ifCLASSINFOpdf
\usepackage[pdftex]{graphicx}
  \graphicspath{{../pdf/}{../jpeg/}}
  \DeclareGraphicsExtensions{.pdf,.jpeg,.png}
\else
  \usepackage[dvips]{graphicx}
  \graphicspath{{../eps/}}
  \DeclareGraphicsExtensions{.eps}
\fi  
\usepackage{amsmath}
\usepackage{cases}
\usepackage{stfloats}
\usepackage{amsfonts}
\usepackage{subeqnarray}
\usepackage{longtable}
\usepackage{supertabular}
\usepackage{setspace}
\usepackage{multirow}
\usepackage{booktabs}
\usepackage[ruled,linesnumbered]{algorithm2e}
\makeatletter
\newcommand{\nosemic}{\renewcommand{\@endalgocfline}{\relax}}
\newcommand{\dosemic}{\renewcommand{\@endalgocfline}{\algocf@endline}}
\let\oldnl\nl
\newcommand{\nonl}{\renewcommand{\nl}{\let\nl\oldnl}}
\makeatother
\usepackage[export]{adjustbox} 
\usepackage{booktabs}
\usepackage{setspace}
\usepackage{xcolor}
\usepackage{mathrsfs}
\usepackage{amsmath}
\usepackage{array}
\usepackage{amssymb}
\usepackage{amsthm}
\usepackage{microtype}
\usepackage{url}
\usepackage{amsfonts,amssymb}
\usepackage{dsfont}
\usepackage{mathtools}
\usepackage{xcolor,colortbl}
\usepackage{colortbl}
\usepackage{graphicx}

\definecolor{Gray}{gray}{0.85}
\definecolor{Whitecolor}{rgb}{1,1,1}

\allowdisplaybreaks
\begin{document}
\setstretch{1}
\title{\textls[-25]{The Design of By-product Hydrogen Supply Chain Considering Large-scale Storage and Chemical Plants: A Game Theory Perspective}}
\author{Qianni~Cao,~\IEEEmembership{Student~Member,~IEEE},
Boda~Li,~\IEEEmembership{Student~Member,~IEEE},
Mengshuo~Jia,~\IEEEmembership{Member,~IEEE}, and 
Chen~Shen,~\IEEEmembership{Senior~Member,~IEEE}

}
        

\maketitle

\begin{abstract}
Hydrogen, an essential resource in the decarbonized economy, is commonly produced as a by-product of chemical plants. To promote the use of by-product hydrogen, this paper proposes a supply chain model among chemical plants, hydrogen-storage salt caverns, and end users, considering time-of-use (TOU) hydrogen price, coalition strategies of suppliers, and road transportation of liquefied and compressed hydrogen. The transport route planning problem among multiple chemical plants is modeled through a cooperative game, while the hydrogen market among the salt cavern and chemical plants is modeled through a Stackelberg game. The equilibrium of the supply chain model gives the transportation and trading strategies of individual stakeholders. Simulation results demonstrate that the proposed method can provide useful insights on by-product hydrogen market design and analysis.
\end{abstract}
\IEEEpeerreviewmaketitle

\section*{Nomenclature}
\addcontentsline{toc}{section}{Nomenclature}

\subsection*{Indices} 
\begin{IEEEdescription}[\IEEEusemathlabelsep\IEEEsetlabelwidth{$aaaaaaaa$}]
	\item[$i,j$]		Index of chemical plants.
	\item[$t$]		Index of time periods during the day.
	\item[$n$]		Index of hydrogen processing equipment, including liquefiers and compressors.
	\item[$I+1$]		Index of the salt cavern.
\end{IEEEdescription}
\subsection*{Parameters} 
\begin{IEEEdescription}[\IEEEusemathlabelsep\IEEEsetlabelwidth{$aaaaaaaa$}]
	\item[$I$]		Number of chemical plants.
	\item[$T$]		Number of time periods.
	\item[$p_{o}$]		Retail price purchased by customers from the salt cavern.
	\item[$\underline{p}_{t},\overline{p}_{t}$]	    Lower and upper bound of the buying price offered by the salt cavern to chemical plants.
	\item[$Q_{trans}$]		Maximal injection rate of the salt cavern.
	\item[$N_\mathcal{C}$]		Number of compressors with different capacity.
	\item[$N_\mathcal{D}$]		Number of liquefiers with different capacity.
	\item[$Q_{i,t}$]		By-product hydrogen quantity produced by chemical plant $i$ in period $t$.
	\item[$\boldsymbol{Q_{pr}}$]        Capacity set of hydrogen processing equipment(kg/h), $\boldsymbol{Q_{pr}}=\{Q_{pr}^{n}\}, \forall n$.
	\item[$Q_\mathcal{C},Q_\mathcal{D}$]        Capacity of a tube trailer and a tanker truck (kg/trip).
	\item[$w_{t}$]      Electricity price in period $t$.
	\item[$\gamma_{c},\gamma_{d}$]      Electricity consumption for unit compressed hydrogen and liquefied hydrogen (kwh/kg).
	\item[$\boldsymbol{K_{1}}$]     Initial investment set of hydrogen processing equipment, $\boldsymbol{K_{1}}=\{K_{1}^{n}\}, \forall n$.
	\item[$K_{2}^{c},K_{2}^{d}$]        Initial investment cost of a tube trailer and a tanker truck.
	\item[$K_{3}$]      Operation cost of a tube trailer (or a tanker truck) in each period.
	\item[$\boldsymbol{T_{a}}$]     $T_{a}^{i,j}$ represents duration from chemical plant $i$ to $j$ ($j= I+1$ represents the salt cavern).
	\item[$\beta_{L1}$]      1 - hourly evaporation rate during the tanker truck loading.
	\item[$\beta_{L2}$]      1 - hourly evaporation rate during transit by a tanker truck.
\end{IEEEdescription}
\subsection*{Decision variables of the salt cavern} 
\begin{IEEEdescription}[\IEEEusemathlabelsep\IEEEsetlabelwidth{$aaaaaaaa$}]
	\item[$p_{t}$]		Buying price the salt cavern offers to chemical plants in period $t$.
	\item[$q_{i,t}^{trans}$]		Hydrogen transaction amount of chemical plant $i$ in period $t$, measured as hydrogen shipped from chemical plant $i$ at the end of the time period $t$.
	\item[$u_{i,I+1}$]		Binary variables. Equals to 1 when products from chemical plant $i$ is shipped directly to the salt cavern. Otherwise, $u_{i,I+1}$ equals to 0.
\end{IEEEdescription}
\subsection*{Decision variables of chemical plants} 
\begin{IEEEdescription}[\IEEEusemathlabelsep\IEEEsetlabelwidth{$aaaaaaaa$}]
	\item[$q_{i,t}^{pr}$]		Hydrogen quantity chemical plant $i$ compressed/liquified in period $t$.
	\item[$\boldsymbol{x_{i}}$]		$\boldsymbol{x_{i}}=\{x_{i}\}, \forall n$ is a set of binary variables. $x_{i}^{n}=1$ when the type of hydrogen processing equipment is selected to purchase. Otherwise, $x_{i}^{n}=0$.
	\item[$N_{i}^{cars}$]		Integer variables of number of tube trailers (or tanker trucks) purchased by chemical plant $i$.
	\item[$u_{i,j}$]		Binary variables. Equals to 1 when products from chemical plant $i$ is shipped to chemical plant $j$. Otherwise, $u_{i,j}$ equals to 0.
	\item[$q_{i,t}^{store}$]		Hydrogen quality in the tube trailer (or tanker truck) left at chemical plant $i$ before filled to capacity in period $t$.
	\item[$q_{i,t}^{unpr}$]		Hydrogen quantity temporarily stored in low-pressure storage tanks before compression or liquefication in period $t$.
	\item[$n_{i,t}^{cars}$]		Integer variables of tube trailers (or tanker trucks) leave chemical plant $i$ in period $t$.
\end{IEEEdescription}

\section{Introduction}
\subsection{Motivation}
\IEEEPARstart{I}{n} 
the context of emission peak and carbon neutrality, hydrogen is not only regarded as a critical alternative to fossil fuel to achieve carbon neutrality but offers versatility and flexibility that renewables cannot reach\cite{Allan2021}. As one of the most cost-effective options, hydrogen produced as a by-product from many chemical plants serves as a cheap and large-scale source of hydrogen. Moreover, by-product hydrogen is usually sufficiently clean and well suited for a wide range of applications, such as fuel cell (FC)-based cogeneration, FC vehicles, domestic heating, and so on\cite{CAMPANARI2020335}. However, the potential of by-product hydrogen has yet to be realized, which is emitted and thus wasted in most cases. Therefore, it presents opportunities as a new revenue stream for chemical plants and promisingly delivers on announced pledges of energy conversion nationwide in the mid-term. However, the lack of infrastructure development such as large-scale storage, logistical supply chain establishment and unexplored market have slowed down its further development.

Salt cavern storage is one of the most promising technologies to achieve large-scale, fast and secure hydrogen storage\cite{ANDERSSON201911901},which offers the most promising option owing to their low investment cost, high sealing potential and low cushion gas requirement\cite{CAGLAYAN20206793}. Notable projects are the salt cavity storages for hydrogen in Teeside, UK, and Texas, USA\cite{Gregoire2019}, demonstrating the operation feasibility on a full industrial scale. However, the business of acquiring, storing and selling by-product hydrogen has not yet been presented as an option by salt cavern operators, which inspires the work to design by-product hydrogen supply chain considering large-scale storage and chemical plants in this paper.

\subsection{Literature Review}

As demand and production capacity for hydrogen grows robustly in recent years, the outlines of hydrogen markets are starting to emerge worldwide. Initial trade and market price discoveries come first on a regional and local basis\cite{James2021}. Infrastructure development, transparent pricing benchmark and logistical supply chain establishment are key growth challenges faced by this new traded commodity just becoming established in energy commodity markets\cite{Allan2021}. 

Presently,  the hydrogen market is far from mature but is showing great potential. Many researchers focus on the planning of the hydrogen supply chain, considering various market scales, hydrogen sources and transportation modes. Life cycle analysis to estimate the economic and environmental benefits was conducted on global\cite{BRANDLE2021117481}, regional\cite{OBARA2019848} or national\cite{REN2020118482} scales. For different hydrogen sources, steam methane reforming (SMR)\cite{CARRERA2021107386} , coal gasification (CG)\cite{LI202027979}, biomass gasification (BG)\cite{CHO2019527,LUMMEN2020118996} and electrolysis (ELE)\cite{WANG2022122194} are common production technologies in recent researches. Considering hydrogen production based on different feedstocks and energy sources, an optimal structure of the hydrogen, biomass and {$\rm CO_{2}$} networks were determined in \cite{GABRIELLI2020115245}. To make comparisons of different transportation modes, Ref. \cite{FAZLIKHALAF202034503} considered four common options with various criteria and scenarios. Ref. \cite{GIM20121162} introduced a method for comparing different transport possibilities of tube or liquid trailer vs. pipeline delivery. The results showed that each transportation technology had a maximally cost-efficient niche and there was no single perfect solution for the entire system. Recently, large-scale storage for liquid hydrogen is of great attention. Ref. \cite{SEO2020114452}  considered  integrated bulk storage of hydrogen and concluded that a centralized storage structure and liquefication in central production plants can reduce the overall cost. Similarly, the status and key gaps for the commercialization of hydrogen liquefication technology with large-scale storage were discussed in \cite{RATNAKAR202124149}. A combination of the hydrogen supply chain with other energy sources has also attracted the attention of many researchers. Ref. \cite{xiao2018} established a local energy market for electricity and hydrogen. Ref. \cite{CARRERA2021116861} proposed a methodological design framework for hydrogen and methane supply chains based on Power-to-Gas systems.

In particular, by-product hydrogen has seen growing attention these years. Ref. \cite{YANEZ2018777} for the first time assessed the economic advantages, the techno-economic feasibility and the central role of reusing by-product hydrogen in the early phase of hydrogen infrastructure in the northern Spain region. A multi-period programming was designed in \cite{YOON2022112083} to make use of existing infrastructure for by-product hydrogen and natural gas (NG) pipelines, which demonstrated the economic benefits of by-product hydrogen. Even though, the potential of by-product hydrogen remains to be discovered.

Meanwhile, most of the literature focuses on maximizing the total benefit of the whole hydrogen supply chain. Ref. \cite{HAN20125328} aimed to maximize social welfare in Korea by planning both capacity and technology of production, storage as well as transportation in an envisioned nationwide hydrogen supply chain. Ref. \cite{WICKHAM2022117740} assessed the effects that hydrogen grades play in the development of a cost-effective hydrogen supply chain. Ref. \cite{EHRENSTEIN2020115486} incorporated the concept of biophysical limits of the planet to address the optimal design of the hydrogen supply chain. An optimization method was proposed in \cite{QUARTON2020113936} for an integrated value chain of carbon dioxide and hydrogen. Individual rationality was introduced in \cite{GUO2021119608}, where the peer-to-peer transaction, endogenous market-clearing price, and uncertainties in hydrogen production were considered in detail. However, most works failed to consider the strategic behaviors and the profit of individual participants, which differed from the usual practice that suppliers and retailers are private companies and operate with a profit-driven mode.

The research gaps for the existing works are: 
\begin{enumerate}
	\item The potential of by-product hydrogen is yet to be realized and its corresponding market is waiting for further exploration.
	\item The dynamic process of chemical plants and salt caverns considering hydrogen generation, compression (or liquefaction), and the transaction is waiting to be modeled.
	\item The interactions and dynamic strategic behaviors of each stakeholder desire a more dedicated modeling framework that captures profits and rationality of individual participants.
\end{enumerate}

\subsection{Contribution}
In this work, we study the by-product hydrogen supply chain considering large-scale storage and multiple chemical plants. The main contributions are threefold:
\begin{enumerate}
	\item We establish a business model for salt caverns to acquire and store by-product hydrogen from chemical plants and sell them to end-users. The by-product hydrogen supply chain composed of each stakeholder in the business model is investigated.
	\item The hour-by-hour decision-making process of each stakeholder, i.e., chemical plants and the salt cavern, is investigated and mathematically modeled under the proposed business model, providing a foundation for the TOU hydrogen pricing strategy. 
	\item The by-product hydrogen market is formulated as a game, considering the individual rationality of each stakeholder. The planning problem among multiple chemical plants is modeled through a cooperative game. The hydrogen market among the salt cavern and chemical plants is modeled through a Stackelberg game, in which the salt cavern is the leader and chemical plants are the followers. 
\end{enumerate}

\section{A business model of salt caverns and chemical plants}
In this section, we develop a business model for salt caverns to acquire by-product hydrogen from chemical plants and sell them to end-users. Generation, large-scale storage, and consumptive way of by-product hydrogen in the business model is introduced first. Then, the comparison between the by-product hydrogen supply chain and the present hydrogen supply chain is made. Followed by this, the structure of the by-product hydrogen market under the proposed business model is introduced in the following section. 

\subsection{Generation, Large-scale Storage and Consumptive Way of By-product Hydrogen}
\begin{figure}[h] 
    \centering
    \includegraphics[width=8.5cm]{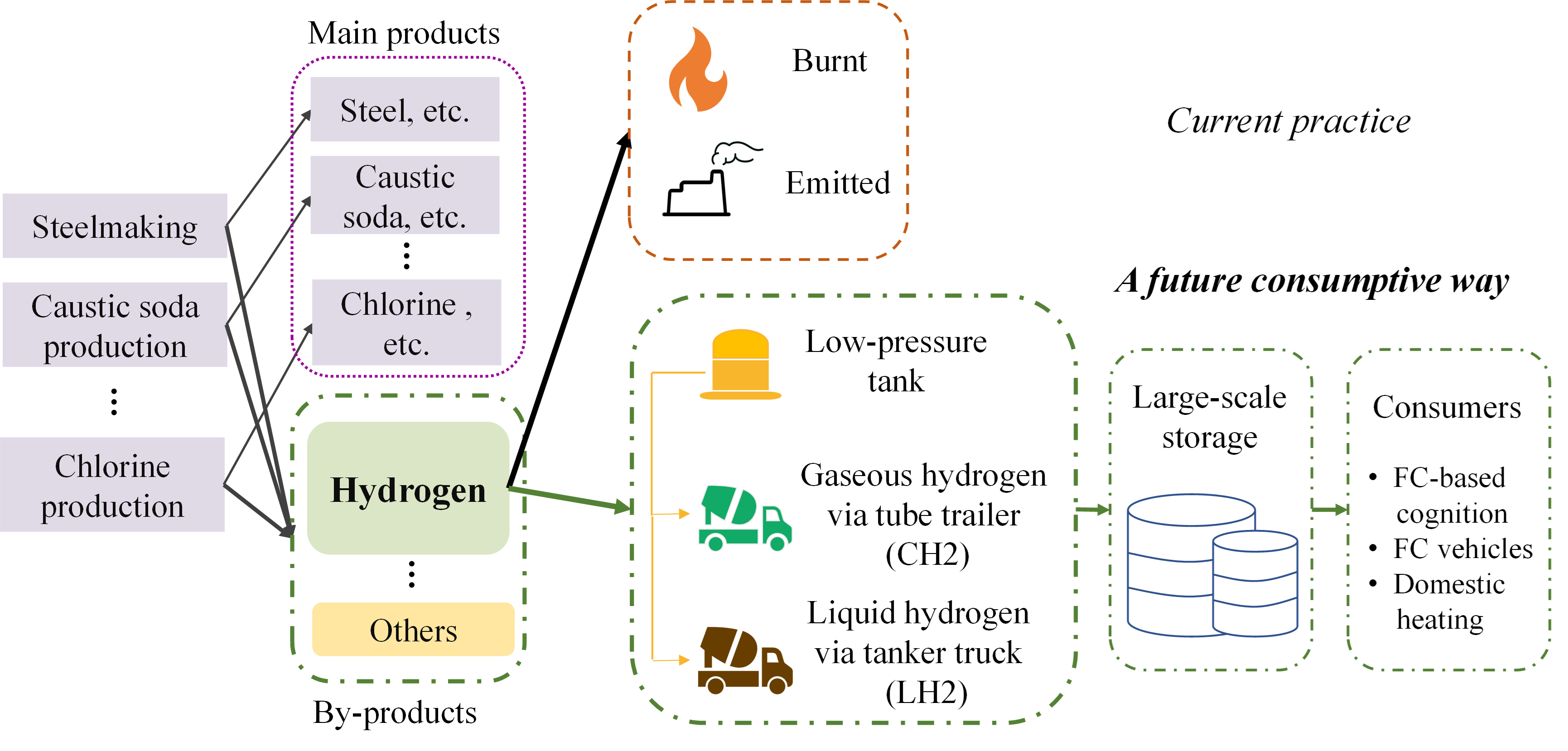}
    \caption{The process of by-product hydrogen generation and its consumptive ways}\label{fig:The process of generation of by-product hydrogen and its consumptive ways}
\end{figure}

By-product hydrogen is a cost-competitive and widely distributed source of hydrogen.
The process of generation of by-product hydrogen and its consumptive ways are illustrated in Fig.\ref{fig:The process of generation of by-product hydrogen and its consumptive ways}.  

Electrochemical processes, such as the industrial production of steel, caustic soda and chlorine, produce hydrogen as a by-product, burnt or emitted as the current practice. However, they can be made available for applications outside chemical plants as a future consumptive way. To transport products from the production facilities to storage sites, by-product hydrogen should be compressed or liquified in advance, which collectively 
are referred to as “hydrogen secondary processing”. Two common transportation modes are compressed gaseous hydrogen via tube trailers (CH2) and liquid hydrogen via tanker trucks (LH2). To alleviate the imbalance between supply and demand of hydrogen, underground cavities like salt caverns are potential to offer natural infrastructure to realize cost-effective and reliable hydrogen storage. At the last link in the supply chain, by-product hydrogen is sold and distributed to various end-users. The proposed generation, storage and consumptive way of hydrogen give rise to a promising by-product hydrogen business model consisting of chemical plants as suppliers, a salt cavern as a retailer and end-users as consumers. 

\subsection{Characteristics of By-product Hydrogen Supply Chain} \label{subsection: Characteristics of by-product hydrogen supply chain}

Differences between the by-product hydrogen supply chain under the proposed business model and most hydrogen supply chains found in literature can be mainly concluded as twofold: 1) composition of major costs; 2) flexibility to coordinate between planning and scheduling. These differences will lead to a distinct focus and a smaller timescale for the formulation of the by-product hydrogen supply chain, which is analyzed as follows:
\begin{table}[h]
\centering
\caption{Major costs of hydrogen supply chain} \label{tab:Major costs of hydrogen supply chain}
\footnotesize
\begin{tabular}{cccc}
\hline\toprule
\multicolumn{2}{c}{\multirow{2}{*}{Major costs}}                                              & \multicolumn{2}{c}{Hydrogen Supply Chain}                        \\ \cline{3-4} 
\multicolumn{2}{c}{}                                                                          & \multicolumn{1}{l}{Traditional} & \multicolumn{1}{l}{By-product} \\ \hline
\multirow{2}{*}{Production}     & Investment & \checkmark &  \\
                                & Operation  & \checkmark &  \\ \hline
\multirow{2}{*}{Storage}        & Investment & \checkmark &  \\
                                & Operation  & \checkmark &  \\ \hline
\multirow{2}{*}{Transportation} & Investment & \checkmark  & \checkmark  \\
                                & Operation  & \checkmark & \checkmark \\ \hline
\multirow{2}{*}{\begin{tabular}[c]{@{}c@{}}Secondary \\ processing\end{tabular}} & Investment &                               & \checkmark                             \\
                                & Operation  &  & \checkmark \\ \hline
\end{tabular}
\end{table}
\begin{table}[h]
\centering
\caption{Major costs and the influence factors}
\footnotesize
\label{tab:Major costs and the influence factors}
\begin{tabular}{cll}
\hline\toprule
\multirow{2}{*}{Major costs} &
  \multicolumn{2}{c}{\multirow{2}{*}{Influence factors}} \\
                                & \multicolumn{2}{c}{}                         \\ \hline
\multirow{2}{*}{Production}     & \multicolumn{2}{l}{1) Production technology} \\
                                & \multicolumn{2}{l}{2) Scale of production}   \\ \hline
\multirow{2}{*}{Storage}        & \multicolumn{2}{l}{1) Storage technology}    \\
                                & \multicolumn{2}{l}{2) Storage capacity}       \\ \hline
\multirow{3}{*}{Transportation} & \multicolumn{2}{l}{1) Transportation mode}   \\
 &
  \multicolumn{2}{l}{\multirow{2}{*}{\begin{tabular}[c]{@{}l@{}}2) Hydrogen volume\\ 3) Transport distance\end{tabular}}} \\
                                & \multicolumn{2}{l}{}                         \\ \hline
\multirow{3}{*}{\begin{tabular}[c]{@{}c@{}}Secondary \\ processing\end{tabular}} &
  \multicolumn{2}{l}{1) Type of processing equipment} \\
 &
  \multicolumn{2}{l}{\multirow{2}{*}{\begin{tabular}[c]{@{}l@{}}2) TOU electricity price\\ 3) Hydrogen volume\end{tabular}}} \\
                                & \multicolumn{2}{l}{}                         \\ \hline
\end{tabular}
\end{table}

\subsubsection{Different composition of major costs}
Major costs of hydrogen supply chain and their influence factors are demonstrated in Table \ref{tab:Major costs of hydrogen supply chain} and \ref{tab:Major costs and the influence factors}, respectively. Unlike the present hydrogen supply chain, producers in the by-product hydrogen supply chain benefit from very low-cost generation. Thus, the major cost comes from secondary processing and transportation.

Power is the major cost for secondary processing. If the liquefier or compressor operates at low-price periods, it may potentially reduce operating costs. Since electricity price fluctuates by hours, the strategic behaviors of each stakeholder should also be modeled by hour.

Transport cost is determined by transportation mode, hydrogen volume and the transport distance. For two transportation modes considered in this paper, LH2 features large transport capacity (often 10-20 times as CH2), high initial investment cost (several times as CH2) and hourly volatile losses. On the contrary, CH2 features low transport capacity, low initial investment cost and zero loss. Usually, for long-distance transportation of a large amount of hydrogen, CH2 is less economical since it requires long rides of much more vehicles than LH2. However, for mid- or short-distance of a small amount of hydrogen, CH2 is more economical since there is no volatile loss. Obviously, a reasonable decision of transportation mode would largely reduce the cost of each chemical plant.
\subsubsection{Less flexibility to coordinate between planning and scheduling }
For suppliers in the by-product hydrogen supply chain, the generation scale of hydrogen is limited by the production plan of their main products. Moreover, their location is less likely to be optimized for the transportation of by-product hydrogen.

\begin{figure}[h] 
    \centering
    \includegraphics[width=8cm]{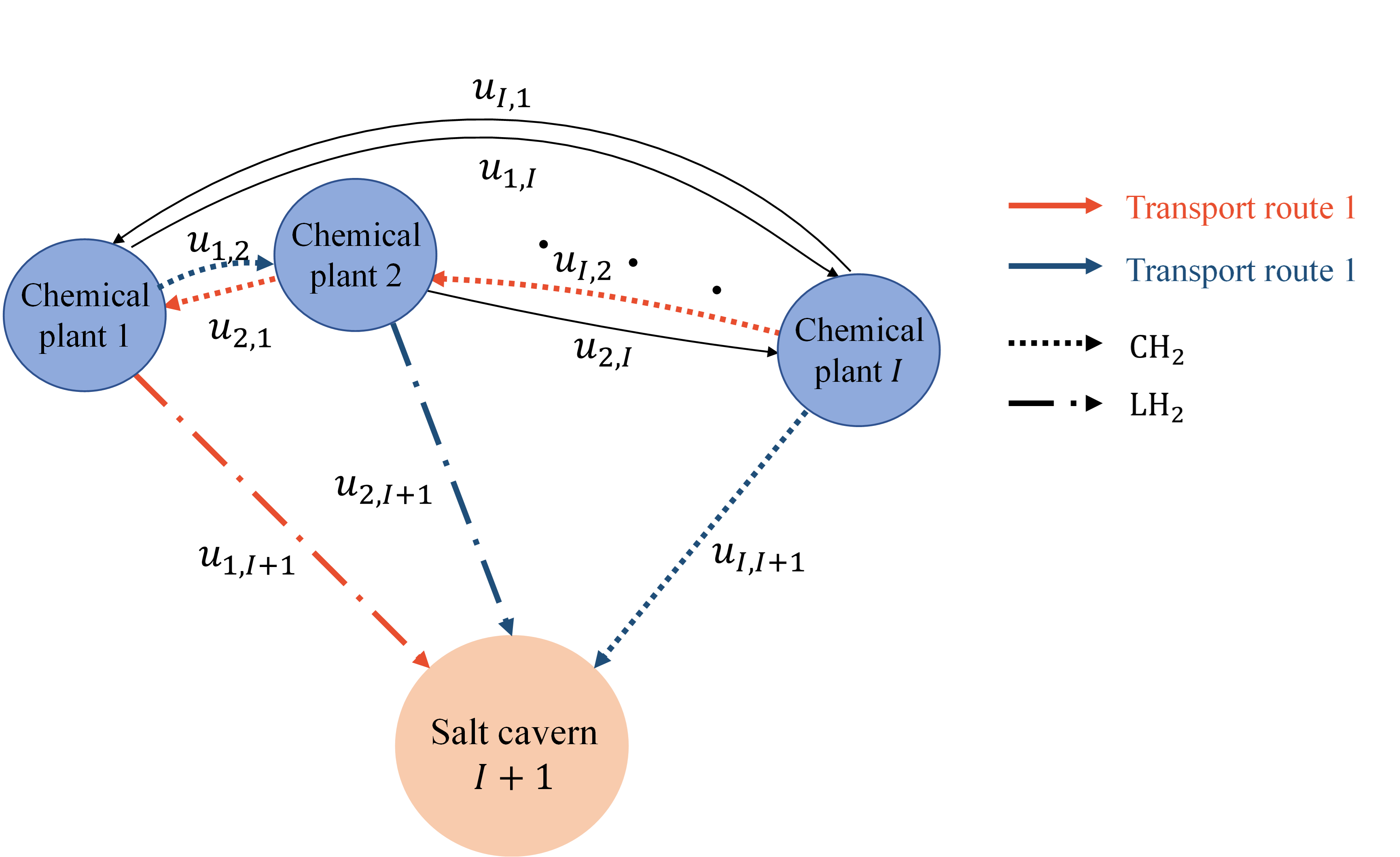}
    \caption{Possible routes for the salt cavern to acquire hydrogen from the chemical plants} \label{fig:Possible routes for the salt cavern to acquire hydrogen from the chemical plants}
\end{figure}

Therefore, there may be a mismatch between each supplier's location and generation scale. Specifically, for distant (to the salt cavern) and medium-yield chemical plants, if CH2 is adopted, long-distance transport of more tube trailers may result in high transportation costs. Nevertheless, if LH2 is adopted, substantial volatile losses would happen due to hours of filling time. This situation results in a dilemma since both transportation mode leads to a revenue decline in some way. Therefore, we envision a scenario where several chemical plants in proximity to each other form a coalition and select a transit hub between them to lower transportation costs, instead of shipping individually to the salt cavern. Two examples of envisioned transportation routes are highlighted in color in Fig.\ref{fig:Possible routes for the salt cavern to acquire hydrogen from the chemical plants}. Moreover, to lower transportation costs, chemical plants destined for the transit hub adopt the CH2 transportation mode, while the transit hub destined for the salt cavern adopt the LH2 transportation mode. In this way, the dilemma between high transportation costs of CH2 and large volatile loss of LH2 is mitigated. 

To sum up, cost structure differences and the lack of flexibility to coordinate between production scale and location lead to a gap between the by-product hydrogen supply chain and the present ones. Therefore, it is essential to model the by-product hydrogen supply chain according to its characteristics rather than simply applying the model of the traditional hydrogen supply chain. 

\subsection{The Structure of By-product Hydrogen Market}

The structure of the proposed by-product hydrogen market is provided in this subsection, followed by the basic assumptions.

\begin{figure}[h] 
    \centering
    \includegraphics[width=8cm]{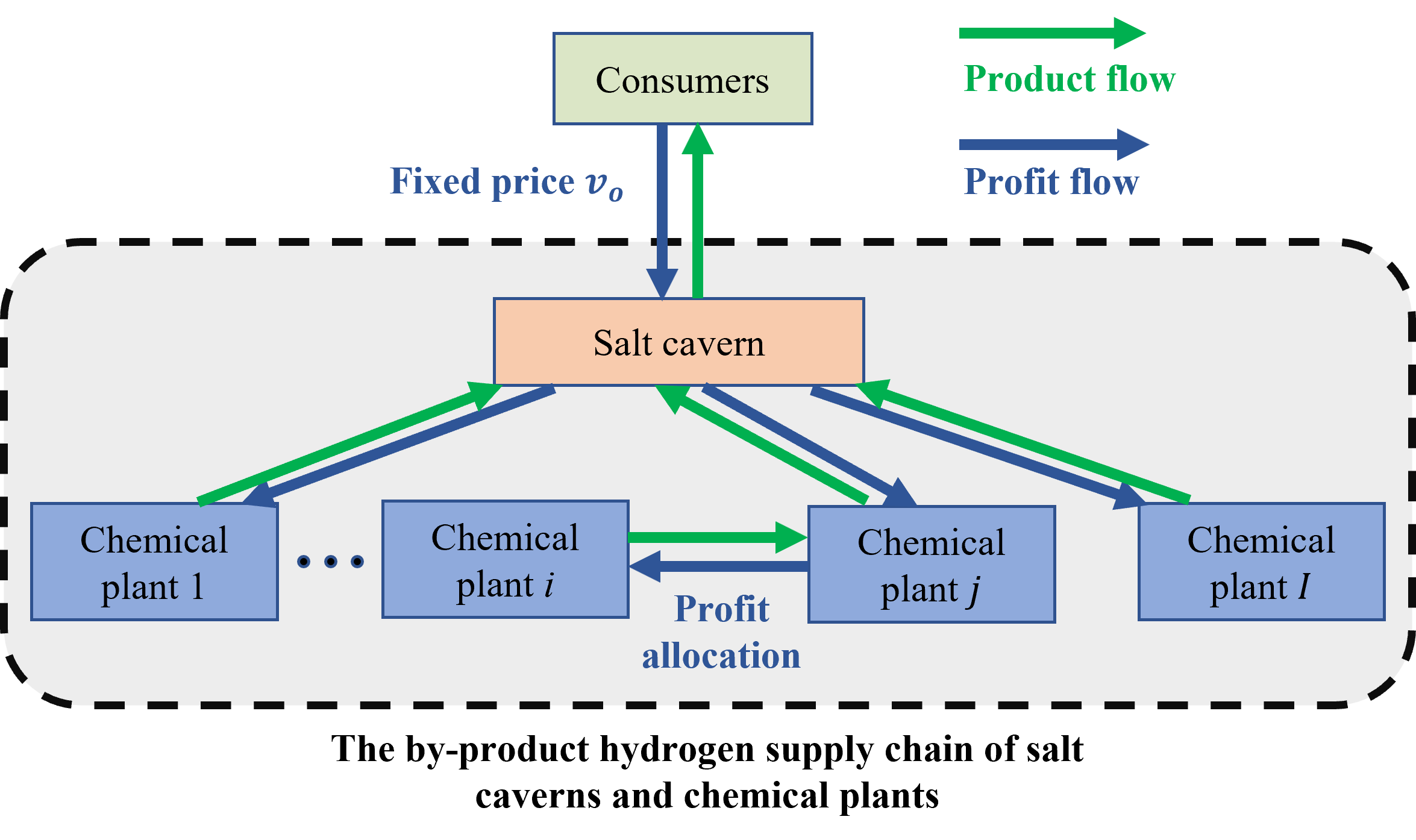}
    \caption{The structure of the by-hydrogen market under investigation} \label{fig:The structure of the by-product hydrogen market under investigation}
\end{figure}
The by-hydrogen market under the proposed business model has the structure illustrated in Fig.\ref{fig:The structure of the by-product hydrogen market under investigation}. Suppliers, namely chemical plants, process by-product hydrogen by liquefiers or compressors (depends on the decision results of each supplier) and deliver it to the retailers. The retailers, namely salt caverns, sell hydrogen to the customers. To simplify the problem, salt caverns are regarded as an entity owned by a single company.

This paper focuses on the transaction between suppliers and retailers. The following assumptions are made without loss of generality:

\begin{enumerate}
	\item The end-users buy all the hydrogen from the retailer at a fixed price. This may happen when the injection-production rate of the salt cavern is higher than the market demand in a region. In order to alleviate the supplier’s market power to drive up prices, we assume that the salt cavern and suppliers have reached such an
    agreement to bring a fixed price into effect. 
	\item The secondary processing cost and transport cost is undertaken by suppliers.  
	\item The production cost is neglected since hydrogen is a by-product of the industrial process of chemical plants. 
	\item Chemical plants would not adjust their production schedule of their main product for the revenue generated by by-product hydrogen.
\end{enumerate}

Based on the above assumptions, the retailer’s and suppliers’ problems can be described as follows. To maximize profits, the salt cavern intends to purchase as much hydrogen as possible from chemical plants at the lowest cost. If the price is too low, chemical plants are less likely to be attracted by this new revenue stream and may waste them as before, which reduces profits of the salt cavern. On the contrary, if the price is too high, the purchasing cost would increase. Therefore, it is important for the salt cavern to strike a balance between the attraction of chemical plants and the purchasing cost. To maximize profits, chemical plants upstream would like to sell more hydrogen when the selling price is high on the one hand, and to reduce processing costs and transport costs on the other hand.

Taking into account the analysis in the last subsection, the challenges of modeling the by-product hydrogen supply chain under the proposed structure are mainly twofold: 1) to explicitly consider possible coalition structures and transport route strategies in the timescale of transport duration, electricity price fluctuation and volatile losses; 2) and to allocate the payoff among the producers in some fairway.

\section{Strategies and decision-making process of stakeholders}

In this section, the decision-making process of each stakeholder is investigated and mathematically modeled under the proposed business model.

\subsection{The Retailer’s Problem}

In the price-setting problem of the salt cavern, the retailer decides its buying price $p_{t}$ (offered to the suppliers), while considering the reactions ${q_{i,t}^{trans}}$ from suppliers. The problem can be formulated as
\setlength{\abovedisplayskip}{3pt}
\begin{align}
    \max\limits_{p_{t}}\ p_{o}\sum_{i=1}^{I}\sum_{t=1}^{T}q_{i,t}^{trans}u_{i,I+1}-\sum_{i=1}^{I}\sum_{t=1}^{T}p_{t}q_{i,t}^{trans}u_{i,I+1}  \label{eq:constraint1}
\end{align}
\begin{align}
    s.t.\ \underline{p}_{t}\le p_{t}\le \overline{p}_{t},\forall t\label{eq:constraint2}
\end{align}
\begin{align}
    \sum\limits_{i=1}^{I}q_{i,t-T_{a}^{i,I+1}}^{trans}\le Q_{trans},\forall t\label{eq:constraint3}
\end{align}

Objective \eqref{eq:constraint1} is the retailer’s profit in which the first term is the selling income, and the second term is the purchasing cost. Inequality \eqref{eq:constraint2} restricts the price offered to suppliers to be within the interval $[\underline{p}_{t},\overline{p}_{t}]$ in each period. Here we assume that the retailer and suppliers have already reached an agreement to bring this constraint into effect. Inequality \eqref{eq:constraint3} prescribes maximal transaction quantity in each period by maximal injection rate of the salt cavern. $q_{i,t}^{trans}$ and $u_{i,I+1}$ are the optimal solution to the suppliers’ problem.

\subsection{The Suppliers’ Problem}

For the suppliers, the planning of the type of processing equipment, transportation mode, and the transport route as well as scheduling of transaction quantity, is formulated in this subsection. To capture the dynamic process of hydrogen transactions between each stakeholder in detail, as well as investigating dynamic strategic behaviors of each stakeholder, the loading process is elaborately taken into consideration. Specifically, hydrogen is produced as a by-product along with main products and has three possible disposal ways: 
\begin{enumerate}
    \item Hydrogen can be loaded to a tube trailer (or a tanker truck) after compression (or liquefaction). At the end of period $t$, tube trailers (or tanker trucks) filled to maximum capacity should depart from chemical plants. Otherwise, they stay until filled up in the following periods. Therefore, the transaction quantity sequence $q_{i,t}^{trans}$ depends on hydrogen processing quantity sequence $q_{i,t}^{pr}$ and capacity of the vehicle ($Q_\mathcal{C}$ for a tube trailer and $Q_\mathcal{D}$ for a tanker truck).
    \item Hydrogen can also be temporarily stored in low-pressure storage tanks before liquefication or achieving an adequate compression rate. It will further be loaded into tube trailers (or tanker trucks) after being compressed (or liquified) in the following periods.
    \item Hydrogen may also be discarded by being emitted or burnt as the current practice, which may happen when buying price offered by the salt cavern is too low or low-pressure storage tanks are filled up.
\end{enumerate}

The above three disposal ways offer multiple options for chemical plants during planning and scheduling. For example, a chemical plant with a generation volume of 100kg per hour, may purchase processing equipment of 100kg per hour. Thus, hydrogen can be processed hour-by-hour. An alternative is to purchase processing equipment of 1000kg per hour. In this case, by-product hydrogen can be temporarily stored in low-pressure storage tanks and will be processed every 10 hours. The suppliers’ problem is to find optimal solutions for planning and scheduling while considering possible coalitions with each other.

In the suppliers’ problem, if the destinations of all suppliers for hydrogen shipment are the salt cavern, decision variables should be the type of processing equipment $\boldsymbol{x_{i}}$ and hydrogen processing amount $q_{i,t}^{pr}$; if the scenario of coalitions of suppliers is taken into account, transport route $u_{i,j}$ of chemical plants $i$ and $j$, which form a coalition.

The decision-making problem, including constraints and objectives of supplier $i$, is given as follows.
\subsubsection{Constraints on transit shipment pattern}
\begin{gather}
    \sum\limits_{j=1}^{I+1}u_{i,j}=1,\forall i\label{eq:constraint4}\\
    u_{i,j}+u_{j,i}\le 1,\forall i,j \label{eq:constraint5}
\end{gather}

Constraint \eqref{eq:constraint4} denotes that the destination of each chemical plant is unique. Constraint \eqref{eq:constraint5} defines that any pairs of the chemical plant $(i,j)$ wouldn’t select each other as the transit destination simultaneously.

\subsubsection{Constraints on hydrogen processing and transport scheduling}

Chemical plants adopting CH2 satisfy: 
\begin{gather}
    n_{i,t}^{cars}\le(q_{i,t}^{pr}+q_{i,t-1}^{store})/Q_c^{car}\le n_{i,t}^{cars}+1,\forall t \label{eq:constraint6}\\
    q_{i,t}^{trans}=n_{i,t}^{cars}Q_\mathcal{C},\forall t \label{eq:constraint7}\\
    q_{i,t}^{store}=q_{i,t-1}^{store}+q_{i,t}^{pr}-q_{i,t}^{trans}, \forall t\in\{2,...T\} \label{eq:constraint8}
\end{gather}

Constraints \eqref{eq:constraint6} and \eqref{eq:constraint7} indicate that hydrogen transaction amount in each period is an integer multiple of the capacity of a tube trailer since only tube trailers filled to maximum capacity will depart from chemical plants. Constraint \eqref{eq:constraint8} denotes variations of hydrogen quantity stored in low-pressure storage tanks.

With the remaining proportion of hydrogen after being shipped from chemical plant $i$ to $j$ ($j= I+1$ represents the salt cavern) written as $\beta_{L2}^{i,j}=\beta_{L2} T_{a}^{i,j}$, chemical plants adopting LH2 satisfy
\begin{gather}
    n_{i,t}^{cars}\le(q_{i,t}^{pr}+\beta_{L1}q_{i,t-1}^{store})/Q_d^{car}\le n_{i,t}^{cars}+1,\forall t \label{eq:constraint9}\\  
    q_{i,t}^{trans}=n_{i,t}^{cars}Q_\mathcal{D}\sum_{j=1}^{I+1}u_{i,j}\beta_{L2}^{i,j},\forall t \label{eq:constraint10}
\end{gather}
\setlength{\abovedisplayskip}{-10pt}
\begin{multline}
    q_{i,t}^{store}=\beta_{L1}q_{i,t-1}^{store}+q_{i,t}^{pr}-{q_{i,t}^{trans}}/{\sum_{j=1}^{I+1}u_{i,j}\beta_{L2}^{i,j}},\\ \forall t \in \{2,...T\} \label{eq:constraint11}
\end{multline}

Constraints \eqref{eq:constraint9} and \eqref{eq:constraint10} indicate that the hydrogen transaction amount in each period is an integer multiple of the capacity of a tanker truck. Constraint \eqref{eq:constraint11} denotes variations of hydrogen quantity stored in low-pressure storage tanks.

Constraints irrelevant to transportation modes are given in \eqref{eq:constraint12}-\eqref{eq:constraint15}, in which the transport duration for chemical plant $i$ is written as $t_{ar}^{i}=\sum_{j=1}^{I+1}u_{i,j}T_{a}^{i,j}$.
\setlength{\abovedisplayskip}{3pt}
\begin{gather}
{\sum_{t - 2\times t_{ar}^{i}}^{t}n_{i,t}^{cars}} \leq N_{i}^{cars}, \forall t \in \left\{2\times t_{ar}^{i},...T \right\} \label{eq:constraint12} \\
q_{i,t}^{pr} \leq {\sum_{n = 1}^{N_{\mathcal{C}} + N_{\mathcal{D}}}x_{i}^{n}}Q_{type}^{n}, \forall t \label{eq:constraint13} \\
q_{i,t}^{unpr} \leq \sum_{n = 1}^{N_{\mathcal{C}} + N_{\mathcal{D}}} x_{i}^{n}Q_{type}^{n}, \forall t \label{eq:constraint14}
\end{gather}
\setlength{\abovedisplayskip}{-3pt}
\begin{multline}
q_{i,t}^{unpr} \leq q_{i,t - 1}^{unprocess} + Q_{i,t} - q_{i,t}^{pr} + {\sum_{j = 1}^{I}{u_{j,i}q_{j,trans}^{t - T_{a}^{i,j}}}},\\\forall t \in \left\{ \max{({1,T_{a}^{i,j}})},\ldots,T \}\right. \label{eq:constraint15} 
\end{multline}
\setlength{\belowdisplayskip}{5pt}

Constraint \eqref{eq:constraint12} imposes the total number of tube trailers (or tanker trucks) purchased by chemical plant $i$ as the upper bound of tube trailers (or tanker trucks) in the round trip during the time period $\left\lbrack t - 2\times t_{ar}^{i} \right\rbrack$. Constraint \eqref{eq:constraint13} prescribes the processing capability of each chemical plant. Constraint \eqref{eq:constraint14} restricts the upper bound of hydrogen stored locally, and the bound parameter is chosen as $\sum_{n=1}^{N_\mathcal{C}+N_\mathcal{D}}x_i^nQ_{type}^n$. Constraint \eqref{eq:constraint15} represents variations of hydrogen stored locally, in which `$\le$' indicates that hydrogen as a by-product can be stored temporarily or directly discarded.

If destinations of all suppliers for hydrogen shipment are the salt cavern, the objective of each chemical plant is to maximize its daily profit and is given in \eqref{eq:constraint16}, in which the income by selling hydrogen to the consumers, initial investment cost and operation cost are considered.
\begin{gather}
\max~\pi_{Fi} = {\sum_{t = 1}^{T}( p_{t}q_{i,t}^{trans} - C_{O}^{i} - C_{T}^{i} )} - C_{INV1}^{i} - C_{INV2}^{i} \label{eq:constraint16} 
\end{gather}
where
\begin{gather}
C_{O}^{i} = q_{i,t}^{pr}{\sum\limits_{n = 1}^{N_{\mathcal{C}} + N_{\mathcal{D}}}x_{i}^{n}}w_{t}\left( \gamma_{c}x_{i}^{c} + \gamma_{d}x_{i}^{d} \right) \label{eq:constraint17}\\
C_{T}^{i} = n_{i,t}^{cars}{\sum_{j = 1}^{I + 1}{u_{i,j}{K_{3}T}_{a}^{i,j}}} \label{eq:constraint18}\\
C_{INV1}^{i} = {\sum_{n = 1}^{N_{\mathcal{C}} + N_{\mathcal{D}}}{x_{i}^{n}K_{1}^{n}}} \label{eq:constraint19}\\
C_{INV2}^{i} = N_{i}^{cars}{({x_{i}^{c}K_{2}^{c} + x_{i}^{d}K_{2}^{d}})} \label{eq:constraint20}
\end{gather}
where transportation mode is written as $x_i^c=\sum_{n=1}^{N_\mathcal{C}}x_i^n$ and $x_i^d=\sum_{n=N_\mathcal{C}}^{N_\mathcal{C}+N_\mathcal{D}}x_i^n$; $C_O^i , C_T^i$ represent hourly processing and transport cost respectively; $C_{INV1}^i$ , $C_{INV2}^i$ represent investment cost of processing equipment and tube trailers (or tanker trucks) after converted into daily cost with a discount rate, respectively.
If the scenario where coalitions of suppliers are considered, we denote chemical plants in a coalition as $\Gamma$. For the chemical plant $i$, $\forall i\in\Gamma$, the objective is to maximize the daily profit of the coalition and is given as \eqref{eq:constraint21}.
\setlength{\belowdisplayskip}{6pt}
\begin{multline}
\max~~\pi_{F\tau}=\sum_{i \in \Gamma}{{\sum_{t = 1}^{T}\left( p_{t}q_{i,t}^{trans}u_{i,I + 1} - C_{O}^{i} - C_{T}^{i} \right)}} \\{ - C_{INV1}^{i} - C_{INV2}^{i}} \label{eq:constraint21}
\end{multline}
where $u_{i,I+1}=1$ when chemical plant $i$ is chosen as a transit hub. Otherwise $u_{i,I+1}=0$.

\section{Game formulation and solution}
\subsection{Game Formulation for By-product Hydrogen Supply Chain}

In this section, the by-product hydrogen market is formulated as a game, considering the individual rationality of each stakeholder. 

The decision-making process of each individual can be concluded as follows. The suppliers plan their initial equipment investment, coalition structure and transport routes in the planning stage. Then, the hydrogen transaction problem, including the retailer's pricing problem and suppliers' scheduling problem, is optimized in the scheduling stage.

\begin{figure}[] 
    \centering
    \includegraphics[width=8.5cm]{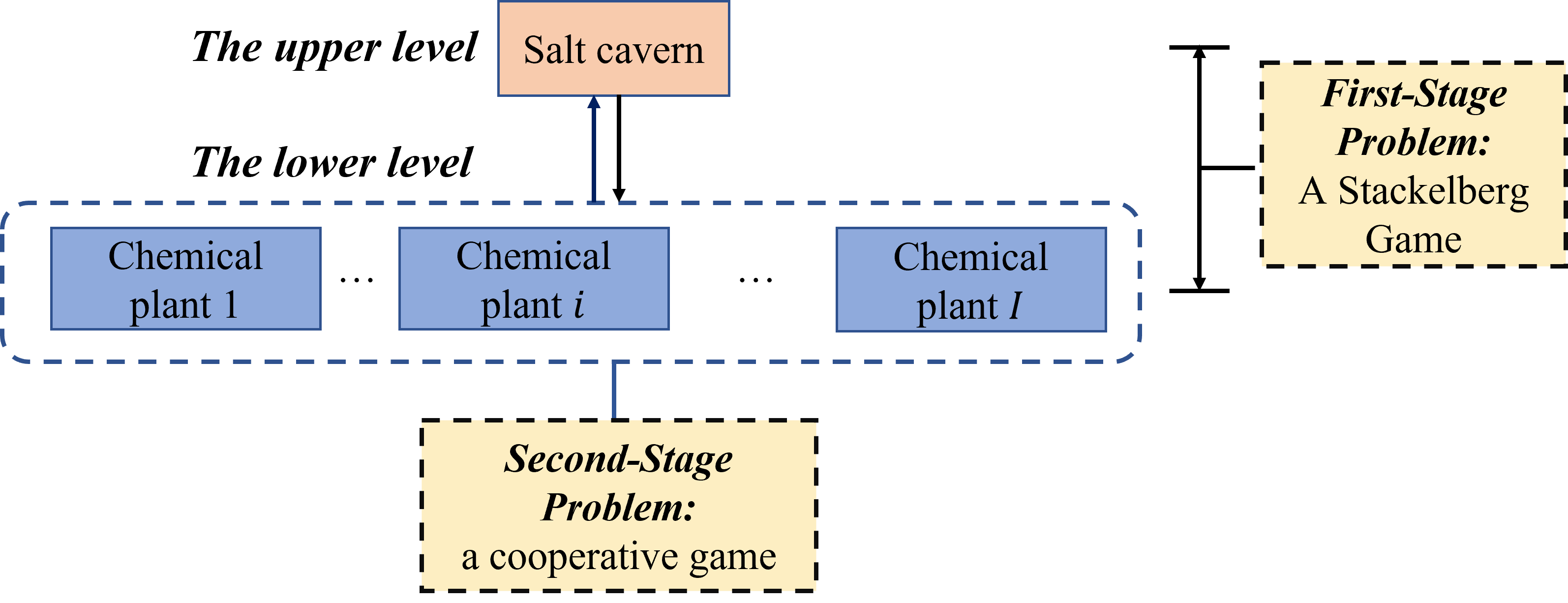}
    \caption{Game models involved in by-product hydrogen supply chain including salt cavern and chemical plants} \label{fig:Game models involved in by-product hydrogen supply chain including salt cavern and chemical plants}
\end{figure}

The overall framework of the game models is illustrated in Fig.\ref{fig:Game models involved in by-product hydrogen supply chain including salt cavern and chemical plants}. Specifically, the planning problem of multiple chemical plants is formulated as a cooperative game, in which a binding coalition could be formed to reduce transport costs. The hydrogen transaction problem between the salt cavern and chemical plants is formulated as a Stackelberg game, in which the salt cavern is the leader and chemical plants are the followers.

\subsubsection{Coorperative game in the planning stage} \label{subsubsection: first-stage problem}
As previously analyzed in subsection \ref{subsection: Characteristics of by-product hydrogen supply chain}, coalitions between chemical plants would potentially lower transportation costs, thus bringing collective payoffs. Moreover, to fairly allocate the payoff $\pi_{F\tau}$ among the players, the Shapley value is adopted.

The following assumptions are made without loss of generality when considering possible coalition structures:

i) Chemical plants in each coalition select one of them as a transit hub to which other chemical plants in the coalition transport hydrogen. Since reducing transport costs is considered as the key factor behind the coalition, we assume that two chemical plants destined for the salt cavern lack the motivation to form a coalition.

ii)	The influence of hydrogen price variations on the coalition structure is neglected since the salt cavern's buying price is unknown at the planning stage. Moreover, the driving force in forming a coalition is to reduce costs rather than to increase the selling income.

Generally, the planning problem of chemical plants is based on the cooperative game, where players are the chemical plants. For chemical plant $i$, decision variables are the type of processing equipment, $\textit{\textbf{x}}_\textit{\textbf{i}}=\left\{x_i^n\right\},\forall n$, hydrogen processing amount $q_{i,t}^{pr}$ and transport route $u_{i,j}$ of chemical plants in the coalition. Payoffs are described as \eqref{eq:constraint16} and \eqref{eq:constraint21} for self-sufficient chemical plants and coalitions respectively.

Note that in the planning stage, the optimal solution $q_{i,t}^{pr}$ is to roughly estimate operation cost under different transportation mode and processing equipment type decisions, thus helping the decision of transport route $u_{i,j}$. Therefore, the solution of $q_{i,t}^{pr}$ here neglects the influence of hydrogen price variations. Actual hydrogen processing quantity sequence $q_{i,t}^{pr}$ will be obtained by equilibrium analysis in the scheduling stage.

\subsubsection{Stackelberg game in the scheduling stage}
The problem in the scheduling is the hydrogen transaction problem between the retailer and the suppliers. After formulating transport route decisions of suppliers as a cooperative game, the interaction between the salt cavern and multiple chemical plants is formulated as a Stackelberg game, where the salt cavern is the leader, whose strategy is the TOU hydrogen price, and chemical plants are followers, whose strategies are hourly transaction. 

At this stage, the retailer’s and suppliers’ problem can be formulated as a bilevel optimization. The retailer determines the hydrogen price sequence $v_t$ in the upper level, and the suppliers decide their optimal transaction pattern $q_{i,t}^{trans}$ in the lower level, with respect to the hydrogen price sequence $v_t$. The optimal transaction pattern $q_{i,t}^{trans}$ would in turn influences hydrogen price sequence $v_t$ determined by the retailer in the upper level. Assume that the information of each chemical plant, such as transit transport routes, processing equipment type and by-product hydrogen generation quantities, are accessible to the salt cavern. Therefore, the optimal solution of $q_{i,t}^{trans}$ can be predicted by the salt cavern under any given hydrogen price sequence $v_t$. The suppliers’ dispatching problem \eqref{eq:constraint3}-\eqref{eq:constraint21} can be regarded as constraints of the retailer’s pricing problem.

According to the analysis above, the interactions between the salt cavern and the chemical plants constitute a Stackelberg competition. In this competition, the salt cavern is the leader, whose strategy is the TOU hydrogen price sequence. Chemical plants are the followers, whose strategy is the hourly hydrogen transaction quantity. The leader’s pricing problem maximizes its profit, subject to the bounds of hydrogen price (Eq.\eqref{eq:constraint2}) and maximal injection rate (Eq.\eqref{eq:constraint3}). The followers’ scheduling problem maximizes individual profits or coalition profits, subject to constraints given in \eqref{eq:constraint9}-\eqref{eq:constraint18}.
\subsection{Solution of the Problem}
In this section, we introduce the solution of the game formulation of the by-product hydrogen supply chain. 

Tractable reformulations of the suppliers’ problem are made to efficiently calculate the equilibrium in the lower level for both the planning and scheduling problems. Specifically, for the suppliers’ problem in both stages, the objective of each individual player (or coalition) is irrelevant to the strategies of other individual players (or coalitions), while the strategy set is influenced by the strategies of other individual players (or coalitions). According to the potential game theory, the suppliers' problem can be regarded as a potential game. The sum of the objectives of each individual player (or coalition) is the potential function. Besides, the pure-strategy equilibrium exists in the transport route planning problem of the suppliers since there exists at least one pure-strategy equilibrium in an infinite potential game. Thus, the suppliers’ problem is formulated as a potential game that can be solved as an optimization problem.

After the reformulation of the suppliers' problem, the planning stage problem is reformulated to a mixed integer nonlinear program (MINLP) with ${\textit{\textbf{x}}_\textit{\textbf{i}},u_{i,j}, q_{i,t}^{pr},N_i^{cars}},\forall i\in{1,\ldots I}$ as decision variables, \eqref{eq:constraint22} as the objective and \eqref{eq:constraint4}-\eqref{eq:constraint15} as constraints. Commercial solvers such as Baron can be used to solve the problem. The solved optimal strategy $\textit{\textbf{x}}_\textit{\textbf{i}}$ and $u_{i,j}$ will be adopted at the scheduling stage.
\setlength{\abovedisplayskip}{3pt}
\begin{multline}
\max~~\pi_{F}~ = ~\sum_{i=1}^{I}{{\sum_{t = 1}^{T}\left( p_{t}q_{i,t}^{trans}u_{i,I + 1} - C_{O}^{i} - C_{T}^{i} \right)}}\\{-C_{INV1}^{i} - C_{INV2}^{i}}  \label{eq:constraint22}
\end{multline}

To solve the bi-level problem at the scheduling stage, Genetic Algorithm (GA) is adopted. First, for the salt cavern in the upper level, pieces of hydrogen price sequences are generated and regarded as individuals. Second, to acquire the fitness of each individual, the suppliers’ scheduling problems in the lower level are solved. Since the transit transport routes $\textit{\textbf{x}}_\textit{\textbf{i}}$ and the processing equipment type $u_{i,j}$ are known at the scheduling stage, the suppliers’ problem becomes a mixed-integer linear program (MILP), which can be solved efficiently by off-the-shelf commercial solvers. Thus, daily profits of the salt cavern, considering the best response of the suppliers, can thus be calculated and regarded as finesses for given price sequences. 

\section{Case Study}
To validate the effectiveness of the proposed model and algorithm, numeric experiments on a by-product hydrogen supply chain composed of three chemical plants and a salt cavern are carried out. All of the following tests are conducted on PCs with Intel Xeon W-2255 processor, 3.70 GHz primary frequency, and 128GB memory. CPLEX 2.16 is used to solve related MILP problems.

\subsection{System Configuration}
Scenario parameters of the envisioned by-product hydrogen supply chain are given in Table \ref{tab:Scenario parameters}. $Q_{i,t}$ are hydrogen generation sequences of a typical day produced by a Gaussian distribution with a mean value of 1000 for the 1st chemical plant (1500 for the 2nd and 3000 for the 3rd) and a variance of 100. Moreover, in the envisioned by-product hydrogen supply chain, $\boldsymbol{Q_{pr}}$ are a vector consisting of 1200, 2000, 4000 and 8000, the first two and the last two of which are the compressor capacity and liquefier capacity to choose from, respectively. Parameters of different processing equipment and transportation modes refer to \cite{HAN20125328} and \cite{Argonne2021} and are given in Table \ref{tab:Parameters of hydrogen transportation}. $\boldsymbol{K_{1}}$ are a vector consisting of 774.29, 126612, 18977.17 and 34757.99, corresponding to each element in $\boldsymbol{Q_{pr}}$. Note that the time scale involved in the problem is one day. Initial investment costs of the liquefier, the compressor, and the transportation vehicles are converted into daily investment costs with a discount rate. The operation cost of a tube trailer (or a tanker truck) in each period includes fuel price, driver wage, and maintenance expenses. 

\begin{table}[h]
\centering
\caption{Scenario parameters of the by-product hydrogen supply chain}
\label{tab:Scenario parameters}
\footnotesize
\begin{tabular}{lllll}
\hline\toprule
\multicolumn{5}{l}{Parameters}                                                           \\ \hline
$I$                      & \multicolumn{2}{l}{3}  & $N_\mathcal{D}$     & 2                             \\
$T$                      & \multicolumn{2}{l}{12} & $T_{a}$     & {[}0,0,0,4;0,0,0,4;0,0,0,4{]} \\
\multicolumn{1}{c}{$p_{o}$} & \multicolumn{2}{l}{15} & $\underline{p}_{t},\overline{p}_{t}$  & 5/13                          \\
$N_\mathcal{C}$                     & \multicolumn{2}{l}{2}  & $Q_{trans}$ & 9000                          \\ \hline
\end{tabular}
\end{table}

\begin{table}[]
\centering
\caption{Parameters of hydrogen transportation} \label{tab:Parameters of hydrogen transportation}
\footnotesize
\begin{tabular}{lllllll}
\hline\toprule
\multicolumn{7}{l}{Parameters}                                                                             \\ \hline
$Q_\mathcal{C}$                        & \multicolumn{4}{l}{200}          & $K_{3}(\$/h)$       & {[}0,0,0,4;0,0,0,4;0,0,0,4{]} \\
$Q_\mathcal{D}$                        & \multicolumn{4}{l}{4000}         & $\beta_{L1}$ & 5/13                          \\
\multicolumn{1}{c}{$\gamma_{c}/\gamma_{d}(kwh/kg)$} & \multicolumn{4}{l}{1/8.18}       & $\beta_{L2}$ & 9000                          \\
$K_{2}^c/K_{2}^d(\$)$                      & \multicolumn{4}{l}{82.20/219.18} &          & \multicolumn{1}{c}{}          \\ \hline
\end{tabular}
\end{table}
\subsubsection{Equilibrium of possible coalition structures of the suppliers}
With three chemical plants, there are five possible coalition structures: no cooperation, cooperation between two players with the third being self-sufficient (there are three ways this could occur) and complete cooperation among all the three chemical plants. The benefits of individual participants or coalitions are shown in Table \ref{tab:Participants/alliance}, in which $M$ represents the benefit, and the benefit of each chemical plant and the sum of them are denoted by $M_{1},M_{2},M_{3}$ and $M_{total}$ respectively. ‘\{\}’ indicates a cooperation, and the chemical plant serving as the transit hub is marked by a ‘*’. 

\begin{table}[]
\centering
\caption{Participants/alliance optimal income under non-cooperative and cooperative game models} \label{tab:Participants/alliance}
\footnotesize
\begin{tabular}{llll}
\hline\toprule
\multirow{2}{*}{Number} &
  \multirow{2}{*}{\begin{tabular}[c]{@{}l@{}}Coalition\\ structure\end{tabular}} &
  \multicolumn{2}{c}{Profits(\$/day)} \\ \cline{3-4} 
  &                & \begin{tabular}[c]{@{}l@{}}Individual or\\ a coalition\end{tabular}        & $M_{total}$ \\ \hline
1 &
  \{1\},\{2\},\{3\} &
  \begin{tabular}[c]{@{}l@{}}$M_{1} = 54052$\\ $M_{2} = 81060$\\ $M_{3} = 236814$\end{tabular} &
  371926 \\
2 & \{1,2*\},\{3\} & \begin{tabular}[c]{@{}l@{}}$M_{\{1,2\}}$ = 170589\\ $M_3$ = 236814\end{tabular} & 407403   \\
3 & \{1,3*\},\{2\} & \begin{tabular}[c]{@{}l@{}}$M_{\{1,3\}}$ = 286531\\ $M_2$ = 107868\end{tabular} & 394399   \\
4 & \{1\},\{2,3*\} & \begin{tabular}[c]{@{}l@{}}$M_1$ = 53562\\ $M_{\{2,3\}}$ = 323154\end{tabular}  & 376716   \\
5 & \{1,2,3*\}     & $M_{\{1,2,3\}}$ = 383925                                                       & 383925   \\ \hline
\end{tabular}
\end{table}

It can be analyzed from Table \ref{tab:Participants/alliance} that:

i)	In the 1st coalition structure with no cooperation at all, the total benefit of the three chemical plants is the lowest among all coalition structures, indicating a potential collective payoff gained by forming coalitions between chemical plants. 

ii)  In the 3rd coalition structure, the benefit of the coalition $\{1, 3^{*}\}$ denoted as $M_{\{1,3^{*}\}}$ equals to 286531 and is lower than the sum of benefits that they could get on their own, which is calculated as $M_{1}+M_{3}=290866$, violating collective rationality.

iii)  In the 5th coalition structure, although collective benefit is higher than the sum of benefits each coalition member could get on their own, the total benefit of the 5th coalition structure $M_{total}\{1,2,3^{*}\}$ is lower than that of the 2nd coalition structure $M_{total}(\{1,2^{*}\},\{3\})$. Therefore, the grand coalition is not stable since there is a preferred alternative. The analysis of the 4th coalition structure is analogous.

iv)  In the 2nd coalition structure, $M_{\{1,2^{*}\}}$, the benefit of the coalition $\{1,2^{*}\}$, equals to 170589 and is higher than the sum of benefits they could get on their own, which satisfies $M_{1}+M_{2}=135112$. Moreover, the total benefit of the 2nd coalition structure is the highest among the five possible structures, so there exists no preferred alternatives. Therefore, the coalition of the chemical plants $\{1,2^{*}\}$ is stable.

The insights provided by different coalition structures above is that for several chemical plants in proximity to each other, those chemical plants with low or medium generation scale (chemical plant 1 and 2 in our case) tends to form a coalition, and to compete with those with larger generation scale.

In order to realize a fair imputation of the collective payoff of chemical plants $\{1,2^{*}\}$, the Shapley value is adopted. The allocation result is {71790.5,98798.5}\$, which is higher than the benefit they could get on their own, which are \{\$54052, \$81060\}. The coalition between chemical plant 1 and 2 increase their profits by 24.7\% and 18.0\% respectively.

\subsection{Equilibrium of Hydrogen Pricing and Scheduling}

In this case, the fixed price at which consumers purchase is set as 15 \$/kg. The equilibrium of the buying price offered by the salt cavern $p_t$ and the hydrogen transaction quantity $q_{i,t}^{pr}$ are illustrated in Fig.5. The minimal price takes value at its lower bound 5\$/kg, and the maximal value is 11.9 \$/kg .
\begin{figure}[] 
    \centering
    \includegraphics[width=7cm]{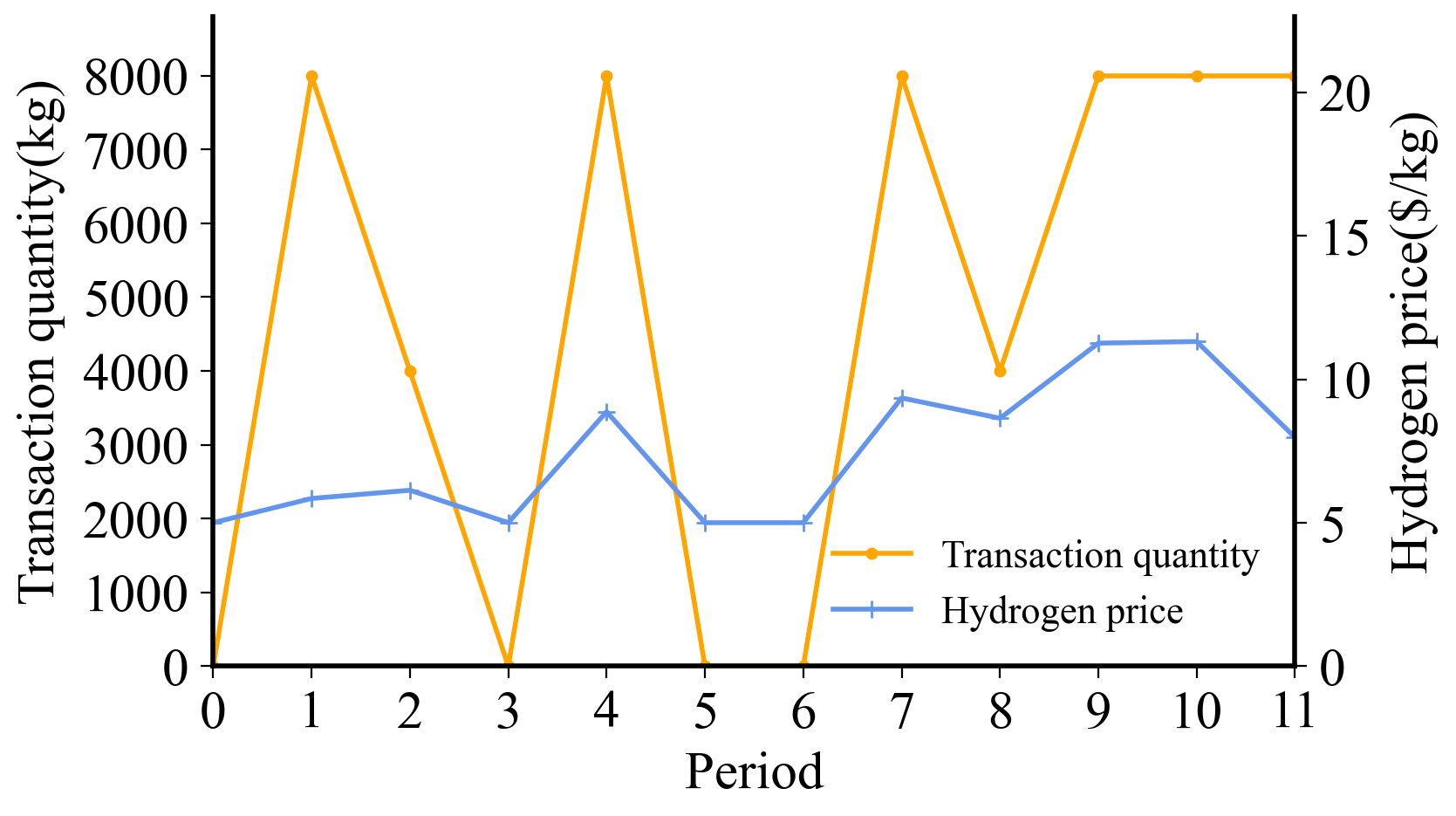}
    \caption{Hydrogen price of salt cavern and transaction quantity of chemical plant} \label{fig:Hydrogen price of salt cavern and transaction quantity of chemical plant}
\end{figure}
It can be observed from Fig.\ref{fig:Hydrogen price of salt cavern and transaction quantity of chemical plant} that the variation trend of the hydrogen transaction quantity goes with the buying price. The higher the buying price, the higher the transaction quantity. This can be attributed to the storage capacity of chemical plants, which can temporarily store by-product hydrogen in low-pressure storage tanks or tube trailers (or tanker trucks) before filled to maximal capacity. Therefore, the chemical plants can choose to sell hydrogen at a higher price.

\begin{figure}[] 
    \centering
    \includegraphics[width=7.5cm]{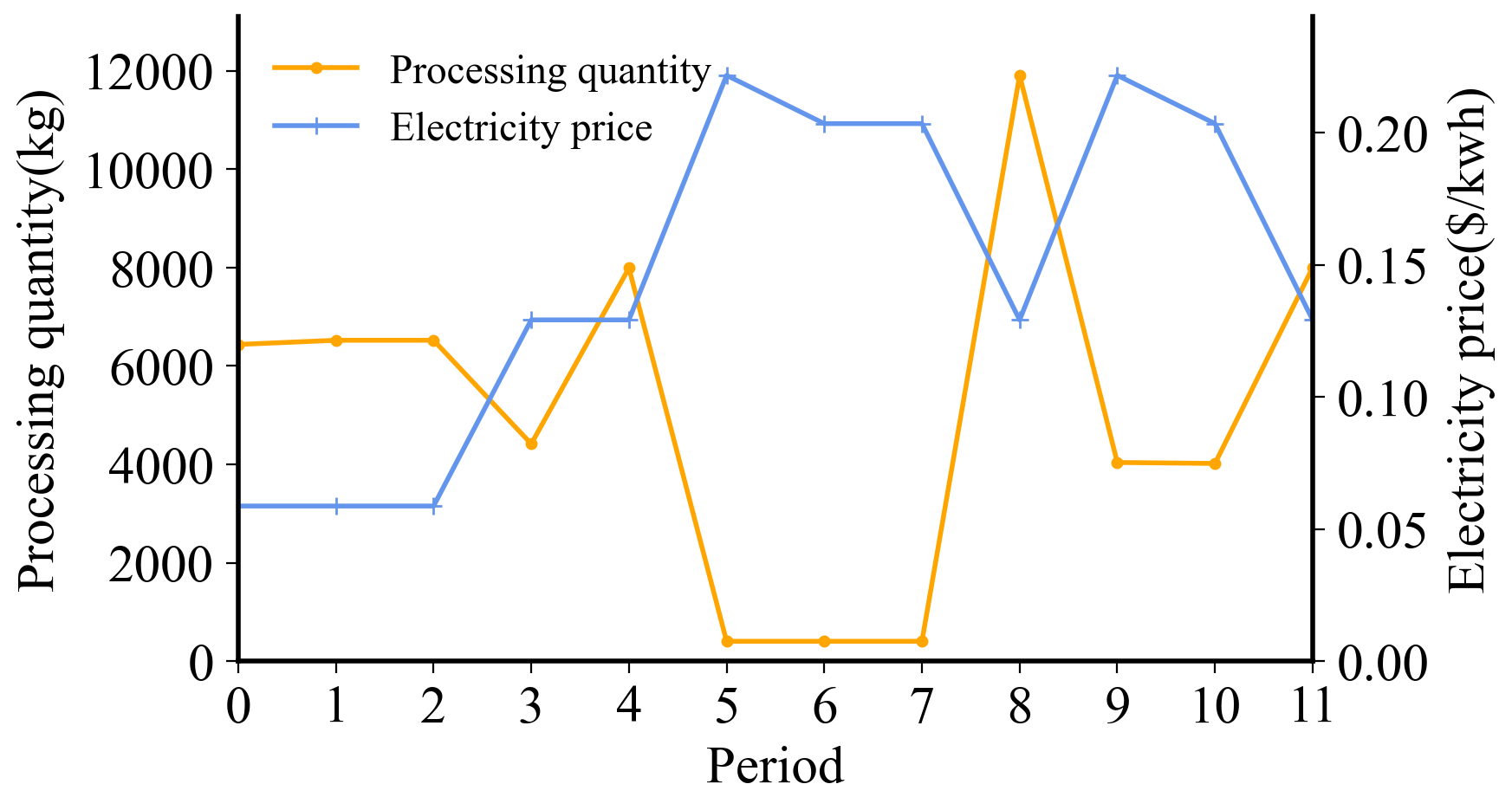}
    \caption{Time of use electricity price and processing mass of chemical plant} \label{fig:Time of use electricity price and processing mass of chemical plant}
\end{figure}

Moreover, due to the influence of the TOU electricity price, the operating cost of the processing equipment fluctuates. The TOU electricity price and the equilibrium of the total processing quantity are plotted in Fig.\ref{fig:Time of use electricity price and processing mass of chemical plant}.

It can be observed from Fig.\ref{fig:Time of use electricity price and processing mass of chemical plant} that the variation trend of the hydrogen processing quantity and the TOU electricity price go oppositely. This is because chemical plants tend to process hydrogen when the electricity price is low, thus reducing the processing cost of hydrogen.

According to the above results, it can be noted that the equilibrium of salt cave pricing encourages chemical plants to process and trade hydrogen when the electricity price is lower. As a result, the salt cavern can purchase hydrogen with lower processing cost, thus reducing the purchase cost of hydrogen per unit. For chemical plants, the hydrogen price is higher during 1-2 periods after periods with lower electricity prices than in other periods, thus reducing the hydrogen processing cost.

\begin{figure}[] 
    \centering
    \includegraphics[width=7.5cm]{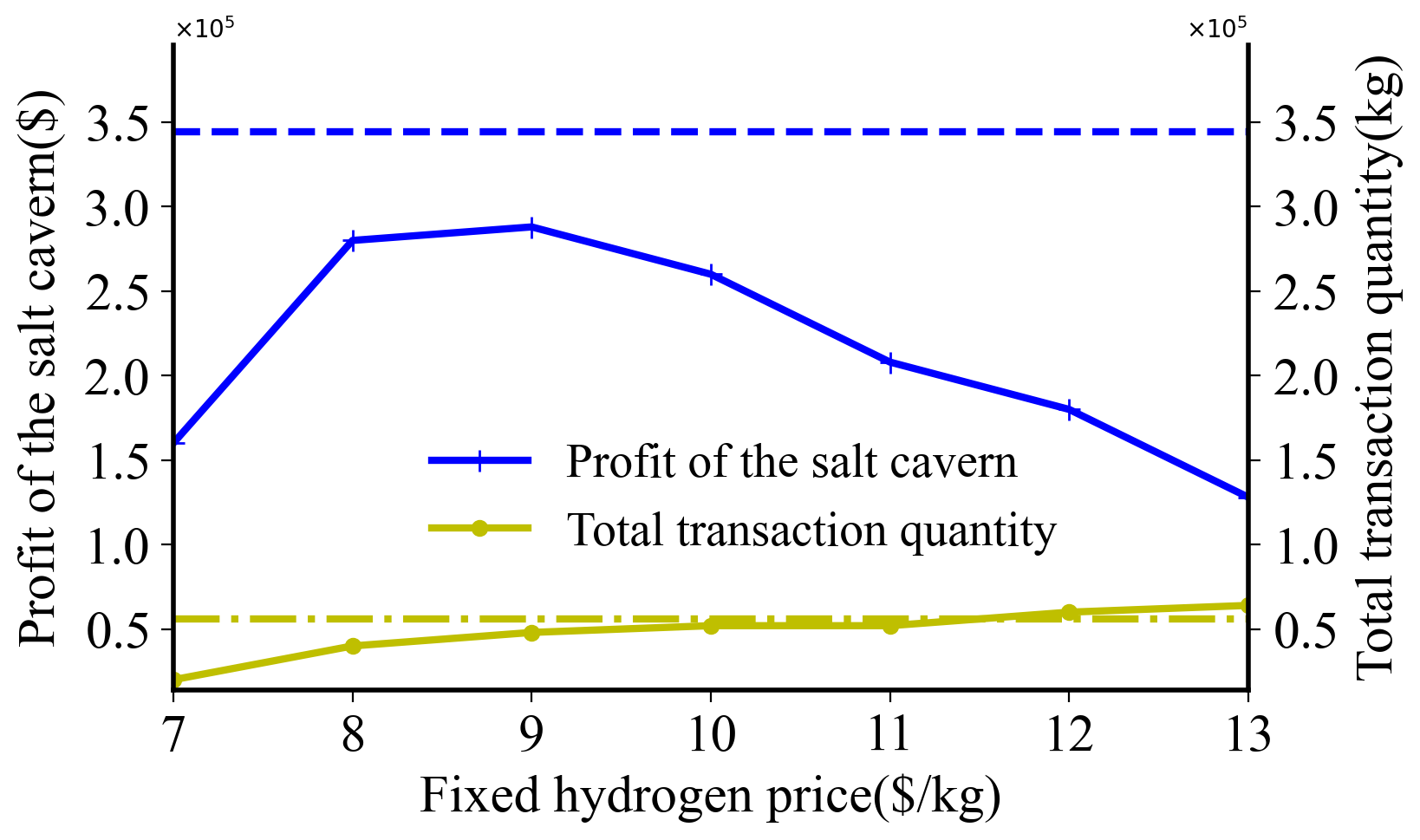}
    \caption{The result of profits and total transaction quantities with time-invariant hydrogen price}\label{fig:The result of profits and total transaction quantities with time-invariant hydrogen price}
\end{figure}

The result of profits and total transaction quantities are plotted in Fig.\ref{fig:The result of profits and total transaction quantities with time-invariant hydrogen price} considering different fixed prices. The optimal price offered by the salt cavern is about 9\$/kg, and its profit is \$287884.8 for a day. However, the profit of the salt cavern reaches to \$343947.16 at the optimal TOU hydrogen price. Hence, a TOU hydrogen price strategy for the salt cavern increases its profit by 19.5\%.

Generally, the equilibrium of the Stackelberg game between the salt cavern and the chemical plants benefits all the players. It also indicates the positive response of the salt cavern and chemical plants to TOU electricity price, and reflects the role of chemical plants in peak shaving and valley filling, which benefits the safe and stable operation of power grid.

\subsection{Sensitivity Analysis}
\subsubsection{Impact of per period transportation operation cost}

\begin{figure}[] 
    \centering
    \includegraphics[width=7cm]{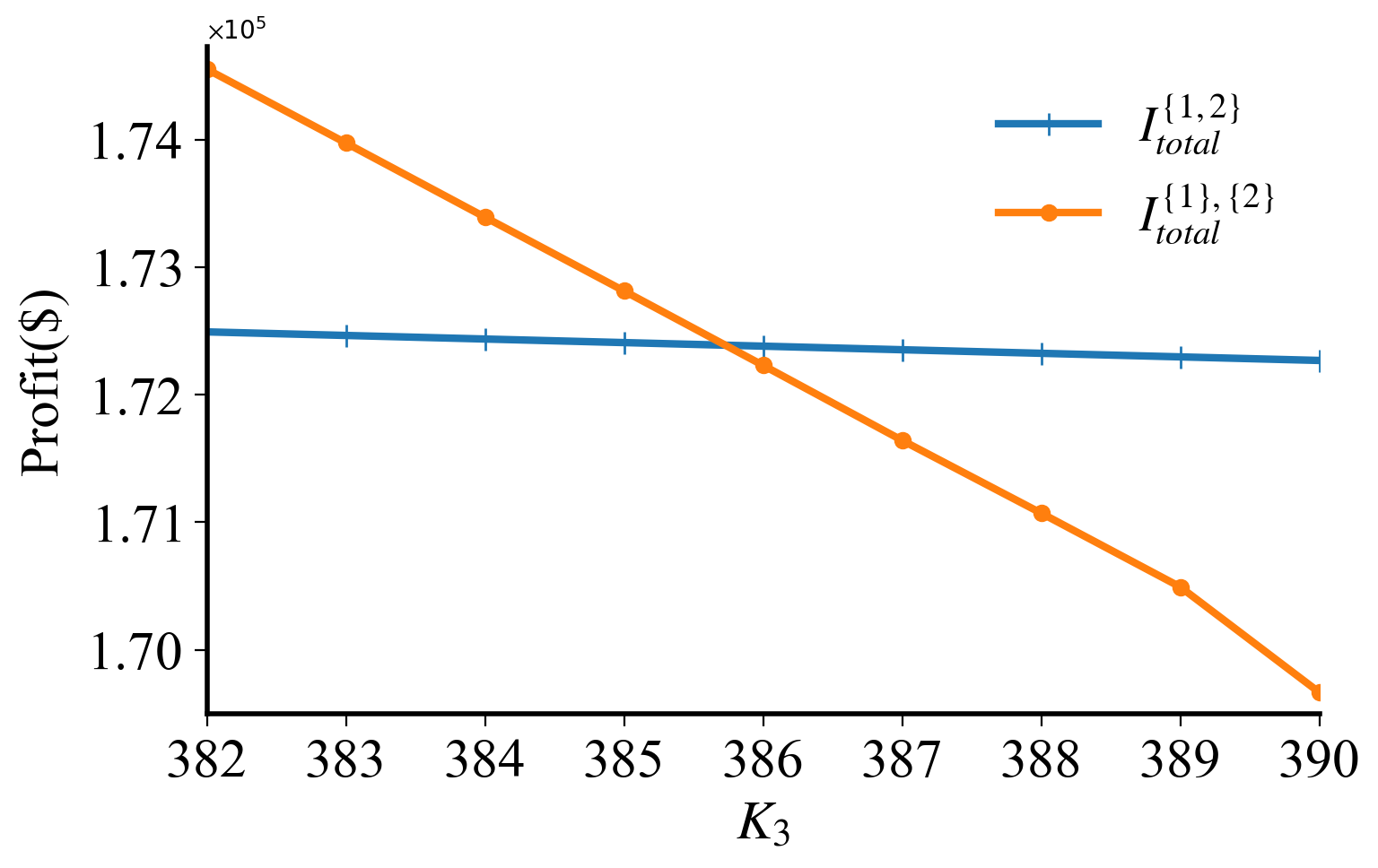}
    \caption{The result of profits and total transaction quantities with time-invariant hydrogen price}\label{fig:Impact of running cost of single vehicle of single period on the profit of chemical plant 1 and 2}
\end{figure}

The reduction in operation cost of a tube trailer (or a tanker truck) per period $K_{3}$ reduces the transport cost, thus bringing down the collective payoff brought by coalitions of chemical plants. Based on the first assumption in section \ref{subsubsection: first-stage problem}, each coalition must take one of them as a transit hub, and two chemical plants destined for the salt cavern lack the motivation to form a coalition. Consequently, the collective payoff declines as the transport cost reduces, until collective rationality no longer holds when the benefits of the coalition are less than the sum of benefits each individual could get on their own. As shown in Fig.\ref{fig:Impact of running cost of single vehicle of single period on the profit of chemical plant 1 and 2}, when $K_{3}$ decreases from \$390 to \$382, the sum of benefits of chemical plant 1 and 2 under the equilibrium of the 1st and 2nd coalition structure, denoted by $M_{total}^{\{1,2\}},M_{total}^{\{1\},\{2\}}$ respectively, gradually increases. 

As illustrated in Fig.\ref{fig:Impact of running cost of single vehicle of single period on the profit of chemical plant 1 and 2}, the coalition benefit is more sensitive to $K_3$ than individual benefits. When $K_3$ decreases to about \$386, the coalition $\{1,2^{*}\}$ no longer bring additional benefits to individuals, resulting in a breakdown of the coalition. 
\subsubsection{Impact of maximal injection rate of the salt cavern}

The maximal injection rate $Q_{trans}$ of the salt cavern directly limits the total transaction quantity per period between the salt cavern and the chemical plants. Table \ref{tab:Individual income} demonstrates the impact of $Q_{trans}$ to the equilibrium of the second-stage problem.

\begin{table}[]
\centering
\caption{Individual income of the equilibrium under different maximum transportation quality of salt cavern gas pipeline in single period} \label{tab:Individual income}
\footnotesize
\begin{tabular}{lllll}
\hline\toprule
\multirow{3}{*}{$Q_{trans}$} &
  \multirow{3}{*}{$M_{total}^{\{1,2\}}$/kg} &
  \multirow{3}{*}{$M_{total}^{\{3\}}$/kg} &
  \multirow{3}{*}{\begin{tabular}[c]{@{}l@{}}$M_{total}$\\(chemical\\ plants)/\$\end{tabular}} &
  \multirow{3}{*}{\begin{tabular}[c]{@{}l@{}}$M_{total}$\\(the salt\\\ cavern)/\$\end{tabular}} \\
      &                              &          &          &           \\
      &                              &          &          &           \\ \hline
12000 & 23103.96                     & 21387.07 & 44491.03 & 342814.83 \\
9000  & \multicolumn{1}{c}{26203.97} & 22762.82 & 48966.80 & 325379.79 \\
6000  & 10528.29                     & 1139.11  & 11667.40 & 278396.59 \\ \hline
\end{tabular}
\end{table}

It can be analyzed from Table \ref{tab:Individual income} that $Q_{trans}$ has different impacts on the participants: the daily income of chemical plants does not necessarily increase with the increase of $Q_{trans}$, whereas the daily income of the salt cavern increases with the increase of $Q_{trans}$. Therefore, the salt cavern will be motivated to determine an appropriate $Q_{trans}$ according to the generation scale of by-product hydrogen of the chemical plants so as to increase individual benefits.

\section{Conclusion}
This paper proposes an equilibrium model of a by-product hydrogen market with the salt cavern as the retailer and chemical plants as the suppliers. A business model for large-scale storage to acquire by-product hydrogen from chemical plants and sell them to end-users is established for the first time. The decision-making process of each stakeholder, i.e., chemical plants and the salt cavern, is investigated and mathematically modeled considering different transportation modes, locations of chemical plants and TOU electricity price. To consider the individual rationality of each stakeholder, the by-product hydrogen market is formulated as games. The transport route planning problem between multiple chemical plants is formulated as a cooperative game. The hydrogen transaction problem between the salt cavern and chemical plants is formulated as a Stackelberg game. Numeric experiments on a by-product hydrogen supply chain composed of three chemical plants and a salt cavern are carried out. The results show that a coalition between chemical plants potentially increases their profits. Moreover, the adoption of TOU hydrogen price in a Stackelberg formulation also increases the profit of the salt cavern. The proposed business model and the optimization of the by-product hydrogen supply chain management not only presents a new revenue stream for both chemical plants and salt caverns but increases resource efficiency and accelerates energy conversion.



\bibliographystyle{IEEEtran}
\bibliography{ref}
\end{document}